\documentclass[aps,twocolumn,a4paper,floatfix,superscriptaddress,pra,longbibliography]{revtex4-2}
\usepackage[toc,page]{appendix}
\usepackage{braket}
\usepackage{epsfig,graphicx,times}
\usepackage{amstext}
\usepackage{amsmath}
\usepackage{amssymb}
\usepackage{mathrsfs}
\usepackage{dcolumn}
\usepackage{bm}
\usepackage[tight]{subfigure}
\usepackage[colorlinks,linkcolor=blue,anchorcolor=blue,citecolor=blue,urlcolor=blue]{hyperref}
\usepackage{color}
\usepackage{siunitx}
\usepackage{appendix}
\usepackage{placeins}
\usepackage{textcomp}
\usepackage{bbold}
\usepackage{float}
\usepackage{verbatim}
\usepackage{minitoc}
\usepackage[dvipsnames]{xcolor}

\usepackage{soul}
\usepackage[framemethod=tikz]{mdframed}
\usepackage{lipsum} 
\usepackage{epstopdf}
\usepackage{CJK}

\begin{document}
	
	\title{Multi-color nonreciprocal optical amplifier with spinning active optomechanics}
	\author{Ru-Ting Sun}
	\affiliation{Key Laboratory of Low-Dimensional Quantum Structures and Quantum Control of Ministry of Education,\\ Department of Physics and Synergetic Innovation Center for Quantum Effects and Applications,\\ Hunan Normal University, Changsha 410081, China}
	
	\author{Mei-Yu Peng}
	\affiliation{Key Laboratory of Low-Dimensional Quantum Structures and Quantum Control of Ministry of Education,\\ Department of Physics and Synergetic Innovation Center for Quantum Effects and Applications,\\ Hunan Normal University, Changsha 410081, China}
	
	\author{Tian-Xiang Lu}
	\email{lu.tianxiang@foxmail.com}
	\affiliation{College of Physics and Electronic Information, Gannan Normal University, Ganzhou 341000, Jiangxi, China}
	
	\author{Ya-Feng Jiao}
	\email{yfjiao@zzuli.edu.cn}
	\affiliation{School of Electronics and Information, Zhengzhou University of Light Industry, Zhengzhou 450001, China}
	\affiliation{Synergetic Innovation Academy for Quantum Science and Technology, Zhengzhou University of Light Industry, Zhengzhou 450002, China}
	
	\author{Jie Wang}
	\affiliation{Key Laboratory of Low-Dimensional Quantum Structures and Quantum Control of Ministry of Education,\\ Department of Physics and Synergetic Innovation Center for Quantum Effects and Applications,\\ Hunan Normal University, Changsha 410081, China}
	
	\author{Qian Zhang}
	\affiliation{Key Laboratory of Low-Dimensional Quantum Structures and Quantum Control of Ministry of Education,\\ Department of Physics and Synergetic Innovation Center for Quantum Effects and Applications,\\ Hunan Normal University, Changsha 410081, China}
	
	\author{Hui Jing}
	\email{jinghui73@foxmail.com}
	\affiliation{Key Laboratory of Low-Dimensional Quantum Structures and Quantum Control of Ministry of Education,\\ Department of Physics and Synergetic Innovation Center for Quantum Effects and Applications,\\ Hunan Normal University, Changsha 410081, China}
	\affiliation{Synergetic Innovation Academy for Quantum Science and Technology, Zhengzhou University of Light Industry, Zhengzhou 450002, China}	
	
	\date{\today}
	
	\begin{abstract}
		We propose to achieve a multi-color nonreciprocal optical amplifier, a crucial device in optical communication and information processing, by spinning an active resonator. We show that in such a device, due to the interplay of the Sagnac effect and the optical gain, nonreciprocal signal {amplification} can be realized, accompanied by a giant enhancement of optical group delay from $0.3\;\mathrm{ms}$ to $35\;\mathrm{ms}$ in a chosen direction, which is otherwise unattainable in a passive device. Also, coherent amplification of higher-order optical sidebands and a slow-to-fast light switch can be achieved by tuning both the pump power and the spinning velocity. Our work provides a unique and accessible way, well-compatible with other existing techniques, to realize multi-color nonreciprocal optical amplifiers for more flexible control of optical fields.
	\end{abstract}
	\maketitle
	
	\section{Introduction}
	
	Nonreciprocal optical devices, featuring distinct properties of light transmission when interchanging the ports of input and output, are indispensable for backscattering-immune optical communications~\cite{kim2002backscattering,fang2017generalized}, optical cloaking~\cite{han2014full,dehmollaian2023transmittable}, and quantum information processing~\cite{sheikh2023asymmetric,scheucher2016quantum,dong2021all,zhang2018thermal,tang2022chiral,tang2022quantum,xia2014reversible,lu2021Nonreciprocity}. Recently, rapid advances have been witnessed in achieving on-chip nonreciprocal devices without any magnetic material, based on space-time modulations~\cite{sounas2017non,li2014photonic,tzuang2014non}, nonlinear and chiral interactions~\cite{qie2023chirality,fan2012all,sang2022spatial,sayrin2015nanophotonic,qie2023chirality,Fan2013Silicon,lodahl2017chiral,nomura2023nonreciprocal,zeng2022nonreciprocal,wu2022passive,xia2018cavity,tang2022nonreciprocal,zhang2018thermal,peng2016chiral}, or non-Hermitian structures~\cite{peng2023nonreciprocal,chen2018nonreciprocity,peng2014parity,shao2020non}. In a very recent experiment, $99.6\%$ optical isolation was achieved by spinning a resonator at $6.6$ kHz~\cite{maayani2018flying}. This provides a versatile platform to explore more new possibilities such as nonreciprocal solitons or chaos~\cite{li2021nonreciprocal,zhang2021nonreciprocal}, nonreciprocal sensing~\cite{jing2018nanoparticle,zhang2020breaking}, and nonreciprocal control of quantum correlations~\cite{jiao2020nonreciprocal,jiao2022nonreciprocal,xiang2023controlling,huang2018nonreciprocal,Li:19,jing2021nonreciprocal,yao2022nonreciprocal,yuan2023optical,jin2021macroscopic,Liu2023NonreciprocalPB}. However, as far as we know, it has not yet been revealed how to achieve multi-color nonreciprocal amplification with such a spinning device, integrated with also an ability to enhance and switch the optical delay or advance.
	
	In parallel, we note that active optical resonators containing gain materials~\cite{krasnok2020active,he2023lasing,liao2023chip,lohof2018prospects,klimov2000optical,gu2023gain,liu2022photonicamplifier,hodaei2014parity,he2013whispering} can radically change optical features or light-matter interactions, thus leading to novel effects such as fast light~\cite{wang2000gain}, inverted optical transparency~\cite{oishi2013inverted,jing2015optomechanically,jiao2016nonlinear}, topological lasing~\cite{Ke2023Topological}, parity-time symmetry breaking~\cite{jing2014pt,he2016dynamical,lu2015p,ozdemir2019parity,zhang2015giant}, and optical nonreciprocity~\cite{peng2014parity,chang2014parity,ma2020chip,zhang2015giant}. Also, in the context of cavity optomechanics (COM), gain-enhanced light-motion couplings have enabled new findings such as thresholdless phonon lasing~\cite{jing2014pt,he2016dynamical,kuang2023nonlinear}, surface-emitting lasing~\cite{yang2015laser,czerniuk2014lasing}, and enhanced COM sensing~\cite{liu2016metrology}. Inspired by these works, we propose to realize a nonreciprocal multi-color optical {amplifier} by spinning an active COM resonator.

	\begin{figure*}[t]
		\centering
		\includegraphics[width=0.9\textwidth]{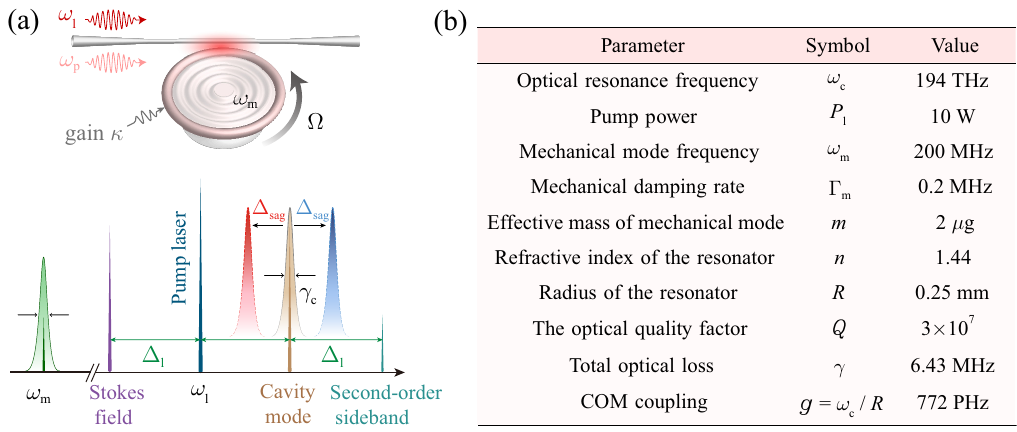}
		\caption{(a) Schematic diagram and frequency spectrum of an active resonator that spins along the CCW direction with an angular velocity $\Omega$. The resonator can support a mechanical mode of frequency $\omega_{\mathrm{m}}$ and is driven by a strong pump laser at frequency $\omega_{\mathrm{l}}$ and a weak probe laser at frequency $\omega_{\mathrm{p}}$. Therefore, we have $\Delta_{\mathrm{sag}}>0$ or $\Delta_{\mathrm{sag}}<0$ for the case with the driving field (including pump laser and probe laser) on the left-hand or right-hand side. (b) Summary of experimentally accessible parameters for numerical simulations~\cite{guo2017universal}.}\label{fig:Schematic diagram}
	\end{figure*}	
	
	Specifically, we consider a spinning COM resonator coupled to a tapered fiber and study the roles of both optical gain and mechanical rotations in the process of optomechanically-induced transparency (OMIT). We show that in this system, multi-color nonreciprocal signal amplification can be realized by tuning the spinning velocity of the resonator, with also well-tunable effects of the nonreciprocal group delay or advance. In addition, nonreciprocal amplification can be realized for higher-order optical sidebands emerging in such an active COM system, indicating the possibility to achieve a nonreciprocal amplifier with a single resonator.
	
	The paper is organized as follows. In Sec.~\ref{Sec:ModelandHamiltonian}, we introduce our model of an active spinning COM system and quantify the optical transmissions. Detailed discussions of the generation and manipulation of the nonreciprocal lights, including those for the probe signal transmission, group delay, and higher-order sidebands, are shown in Sec.~\ref{Sec: Results and discussions}, and a summary is given in Sec.~\ref{Sec:Conclusions}.
	
	\section{Theoretical model}\label{Sec:ModelandHamiltonian}
	
	As shown in Fig.~\ref{fig:Schematic diagram}(a), we consider a spinning whispering-gallery mode resonator with frequency $\omega_{\mathrm{c}}$ and the optical intrinsic decay rate $\gamma_{\mathrm{c}}=\omega_{\mathrm{c}}/Q$ (with $Q$ being the optical quality factor). This resonator supports a mechanical radial breathing mode (with frequency $\omega_{\mathrm{m}}$ and effective mass $m$) that is coupled to the optical mode via optical-radiation-pressure induced interaction. To explore the OMIT effect, a strong pump laser is applied at frequency $\omega_{\mathrm{l}}$ and a weak probe laser at frequency $\omega_{\mathrm{p}}$ to drive the system, with the field amplitudes $$\varepsilon_{\mathrm{l}}=\sqrt{\gamma_{\mathrm{ex}}P_{\mathrm{l}}/\hbar\omega_{\mathrm{l}}},~~~
	\varepsilon_{\mathrm{p}}=\sqrt{\gamma_{\mathrm{ex}}P_{\mathrm{s}}/\hbar\omega_{\mathrm{p}}},$$ respectively, with $P_{\mathrm{l}}$ ($P_{\mathrm{s}}$) the input power of the pump (probe) laser and $\gamma_{\mathrm{ex}}$ the external decay rate of resonator-fiber coupling~\cite{safavi2011electromagnetically,weis2010optomechanically}. Hereafter, we denote the total optical loss rate by $\gamma=\left(\gamma_{\mathrm{c}}+\gamma_{\mathrm{ex}}\right)/2$.
	
	Our aim here is to study how to manipulate the OMIT effect by combining the roles of the Sagnac effect and the gain. We note that in experiments, the active resonator can be fabricated by using gain materials~\cite{peng2014parity,shao2020non}, such as a silica microtoroid doped with Er$^{3+}$ ions which can emit photons in the $1550\,\textrm{nm}$ band when driven by a laser in the $980\,\textrm{nm}$ or $1450\,\textrm{nm}$ band~\cite{peng2014parity}. Also, very recently, the rotary resonator was realized by mounting on a turbine spinning along its axis~\cite{maayani2018flying}. For the resonator with radius $R$ and angular velocity $\Omega$, the frequency of the optical mode undergoes an opposite Sagac-Fizeau frequency shift~\cite{malykin2000sagnac,mao2022experimental,yang2022nonreciprocal,Burns2022EngineeringOI,jiang2018nonreciprocal,shi2019gauge}, i.e., $\omega_{\mathrm{c}}\rightarrow\omega_{\mathrm{c}}+\Delta_{\mathrm{sag}}$, with
	\begin{align}	\Delta_{\mathrm{sag}}=\pm\Omega\frac{nR\omega_{\mathrm{c}}}{c}\left(1-\frac{1}{n^{2}}-\frac{\lambda}{n}\frac{dn}{d\lambda}\right),
	\end{align}
	where $n$ is the refractive index of the resonator and $c~(\lambda)$ is the speed (wavelength) of light in vacuum. The dispersion term $dn/d\lambda$, describing the relativistic origin of the Sagnac effect, is relatively small in typical materials (reaching up to $\sim1\% $)~\cite{malykin2000sagnac,maayani2018flying}. Fixing the counterclockwise (CCW) rotation of the resonator, we have $\Delta_{\mathrm{sag}}>0$ ($\Delta_{\mathrm{sag}}<0$) for the case with the driving field (including pump laser and probe laser) input from the left-hand (right-hand) side.	
	
	The effective Hamiltonian of the system, in a frame rotating at driving frequency $\omega_{\mathrm{l}}$, can be written as
	\begin{align}\label{H}
		H	&=\hbar\left(\Delta_{\mathrm{l}}+\Delta_{\mathrm{sag}}\right)a^{\dagger}a+\frac{p^{2}}{2m}+\frac{1}{2}m\ensuremath{\omega_{\mathrm{m}}^{2}}x^{2}+\frac{p_{\theta}^{2}}{2mR^{2}}\nonumber\\
		&\quad-\hbar ga^{\dagger}ax+i\hbar\left(\varepsilon_{\mathrm{l}}a^{\dagger}+\varepsilon_{\mathrm{p}}a^{\dagger}e^{-i\xi t}-\mathrm{H.c.}\right),
	\end{align}where $\Delta_{\mathrm{l}}=\omega_{\mathrm{c}}-\omega_{\mathrm{l}}$, $\xi=\omega_{\mathrm{p}}-\omega_{\mathrm{l}}$, and $g=\omega_{\mathrm{c}}/R$ denotes the COM coupling rate. $a\,(a^{\dagger})$ is the optical annihilation (creation) operator, and $x$, $\text{\ensuremath{p}}$, $\theta$ and $p_{\theta}$ are displacement, momentum, rotation angle and angular momentum operators, respectively, which satisfy the commutation relations $\left[x,p\right]=\left[\theta,p_{\theta}\right]=i\hbar$~\cite{davuluri2016controlling}.
	
	The mean response of the system to the probe signal, rather than including quantum fluctuation, is processed by OMIT in COM systems~\cite{safavi2011electromagnetically,weis2010optomechanically}. As a result, for the sake of exploring the nonlinear dynamics of it, the equations of motion of the system can be derived as
	\begin{equation} \label{EOM}
		\begin{aligned}
			&\dot{a}= \left[-i\left(\Delta_{\mathrm{l}}+\Delta_{\mathrm{sag}}\right)+\kappa+\ensuremath{igx}\right]a+\varepsilon_{\mathrm{l}}+\varepsilon_{\mathrm{p}}e^{-i\xi t},\\
			&\ddot{x}=-\text{\ensuremath{\omega_{\mathrm{m}}^{2}}}x+\frac{\hbar g}{m}a^{\dagger}a-\Gamma_{\mathrm{m}}\dot{x}+\frac{p_{\theta}^{2}}{m^{2}R^{3}},\\
			&\dot{\theta}=\frac{p_{\theta}}{mR^{2}},\\
			&\dot{p}_{\theta}=~0.\\
		\end{aligned}
	\end{equation}Note that the optical or mechanical damping terms are added phenomenologically into the above equation. In the following, we have assumed that for $\kappa/\gamma=-1$, $\kappa$ denotes the optical decay, whereas for $\kappa/\gamma=1$, $\kappa$ denotes the optical gain. The steady-state values of the dynamical variables are obtained:
	\begin{equation}\label{the steady-state solutions}
		\begin{aligned}
			a_{\mathrm{s}}=~ \frac{\varepsilon_{\mathrm{l}}}{i\Delta-\kappa},~~x_{\mathrm{s}}=~\frac{\hbar g}{m\ensuremath{\omega_{\mathrm{m}}^{2}}}\left|a_{\mathrm{s}}\right|^{2}+R\left(\frac{\Omega}{\ensuremath{\omega_{\mathrm{m}}}}\right)^{2},\\
		\end{aligned}
	\end{equation}
	where $\Delta=\Delta_{\mathrm{l}}+\Delta_{\mathrm{sag}}-gx_{\mathrm{s}}$ is the effective optical detuning. Obviously, the steady-state mechanical displacement $x_{\mathrm{s}}$ and the intracavity optical amplitude $a_{\mathrm{s}}$ are determined by the angular velocity $\Omega$ and the optical gain $\kappa$. As a result, the effective COM coupling rate and the breathing mode oscillations of the active spinning COM system can be tuned by adjusting the rotating speed and the gain-to-loss ratio, which in turn leads to the modified OMIT properties of the system.
	
	For the scenario where the probe light for OMIT process is generally much weaker than the pump light~\cite{lu2023magnon}, every operator can be expanded as a sum of its steady-state mean value and small fluctuations around that value, that is,	\begin{equation}
		\begin{aligned}
			x=x_{\mathrm{s}}+\delta x,
			\,a=a_{\mathrm{s}}+\delta a.
		\end{aligned}
	\end{equation}
	The pump light supplies the steady-state solution of the optomechanical system, while the probe light is regarded as the perturbation of the steady-state~\cite{safavi2011electromagnetically,weis2010optomechanically}. Accordingly, the fluctuation terms satisfy the following equations\begin{equation}
		\begin{aligned}\label{EQ6}
			\delta\dot{a}&=\left(-i\Delta+\kappa\right)\delta a+\ensuremath{ig\left(a_{\mathrm{s}}\delta x+\delta a\delta x\right)}+\varepsilon_{\mathrm{p}}e^{-i\xi t},\\
			\delta\ddot{x}&=-\text{\ensuremath{\omega_{\mathrm{m}}^{2}}}\delta x-\Gamma_{\mathrm{m}}\delta\dot{x}+\frac{\hbar g}{m}\left(a_{\mathrm{s}}\delta a^{\dagger}+a_{\mathrm{s}}^{*}\delta a+\delta a^{\dagger}\delta a\right).
		\end{aligned}
	\end{equation}
	To calculate the amplitudes of the first- and second-order sidebands, we solve these equations by using the ansatz ~\cite{weis2010optomechanically}:
	\begin{equation}
		\begin{aligned} \label{ansatz}
			\delta x=&X_{(1)}e^{-i\xi t}+X_{(1)}^*e^{i\xi t}+X_{(2)}e^{-2i\xi t}+X_{(2)}^*e^{2i\xi t},\\
			\delta a=&A^{+}_{(1)}e^{-i\xi t}+A^{-}_{(1)}e^{i\xi t}+A^{+}_{(2)}e^{-2i\xi t}+A^{-}_{(2)}e^{2i\xi t},
		\end{aligned}
	\end{equation}
	where the plus and minus signs correspond to the upper and lower sidebands~\cite{zhang2018loss}. By substituting Eq.~(\ref{ansatz}) into Eq.~(\ref{EQ6}) and neglecting the second-order parts, the linear response of the system can be obtained as
	\begin{equation}
		\begin{aligned}\label{Solutions}
			A^{+}_{(1)}=\frac{\varepsilon_{\mathrm{p}}\left[\lambda_{-}(\xi)\lambda(\xi)+i\hbar g^{2}\left|a_{\mathrm{s}}\right|^{2}\right]}{\lambda_{+}(\xi)\lambda_{-}(\xi)\lambda(\xi)-2\hbar \Delta g^{2}\left|a_{\mathrm{s}}\right|^{2}},\\
			X_{(1)}=\frac{\lambda_{-}(\xi)\hbar ga_{\mathrm{s}}^{*}\varepsilon_{\mathrm{p}}}{\lambda_{+}(\xi)\lambda_{-}(\xi)\lambda(\xi)-2\hbar \Delta g^{2}\left|a_{\mathrm{s}}\right|^{2}},
		\end{aligned}
	\end{equation}
	where
	\begin{equation*}
		\begin{aligned}
			\lambda(\xi)=~m\left(-\xi^{2}+\text{\ensuremath{\omega_{\mathrm{m}}^{2}}}-i\xi\Gamma_{\mathrm{m}}\right),~~
			\lambda_{\pm}(\xi)=-i\xi-\kappa\pm i\Delta.
		\end{aligned}
	\end{equation*}
	By combining Eqs.~(\ref{EQ6})-(\ref{Solutions}), the second-order sideband amplitude is given by
	\begin{align}
		A^{+}_{(2)}=\frac{C_{1}X_{(1)}^2+C_{2}A^{+}_{(1)}X_{(1)}}{\lambda_{-}(\xi)C_{3}},
	\end{align}
	where
	\begin{align*}
		\begin{aligned}
			C_{1}=&-i\hbar g^{4}\left|a_{\mathrm{s}}\right|^{2}a_{\mathrm{s}},\\
			C_{2}=&~i g\lambda_{-}(\xi)\lambda(2\xi)\lambda_{-}(2\xi)-i\hbar \xi g^{3}\left|a_{\mathrm{s}}\right|^{2},\\
			C_{3}=&~\lambda(2\xi)\lambda_{+}(2\xi)\lambda_{-}(2\xi)-2\hbar \Delta g^{2}\left|a_{\mathrm{s}}\right|^{2}.
		\end{aligned}
	\end{align*}
	
	Using the standard input-output relation~\cite{gardiner1985input}: $a_{\mathrm{out}}=a_{\mathrm{in}}-\sqrt{\gamma_{\mathrm{ex}}}a$, the transmission rate of the probe light~\cite{jing2015optomechanically} and the efficiency of the second-order upper sideband~\cite{jiao2016nonlinear} will be obtained as
	\begin{equation}
		\begin{aligned}
			T=&\left|t_{\mathrm{p}}\right|^{2}=\left|\frac{a_{\mathrm{out}}}{a_{\mathrm{in}}}\right|^{2}=\left|1-\frac{\gamma_{\mathrm{ex}}}{\varepsilon_{\mathrm{p}}}A^{+}_{(1)}\right|^{2},\\\eta=&\left|\frac{\gamma_{\mathrm{ex}}}{\varepsilon_{\mathrm{p}}}A^{+}_{(2)}\right|.
		\end{aligned}
	\end{equation}
\begin{figure}[htbp]
		\centering
		\includegraphics[width=\linewidth,scale=0.95]{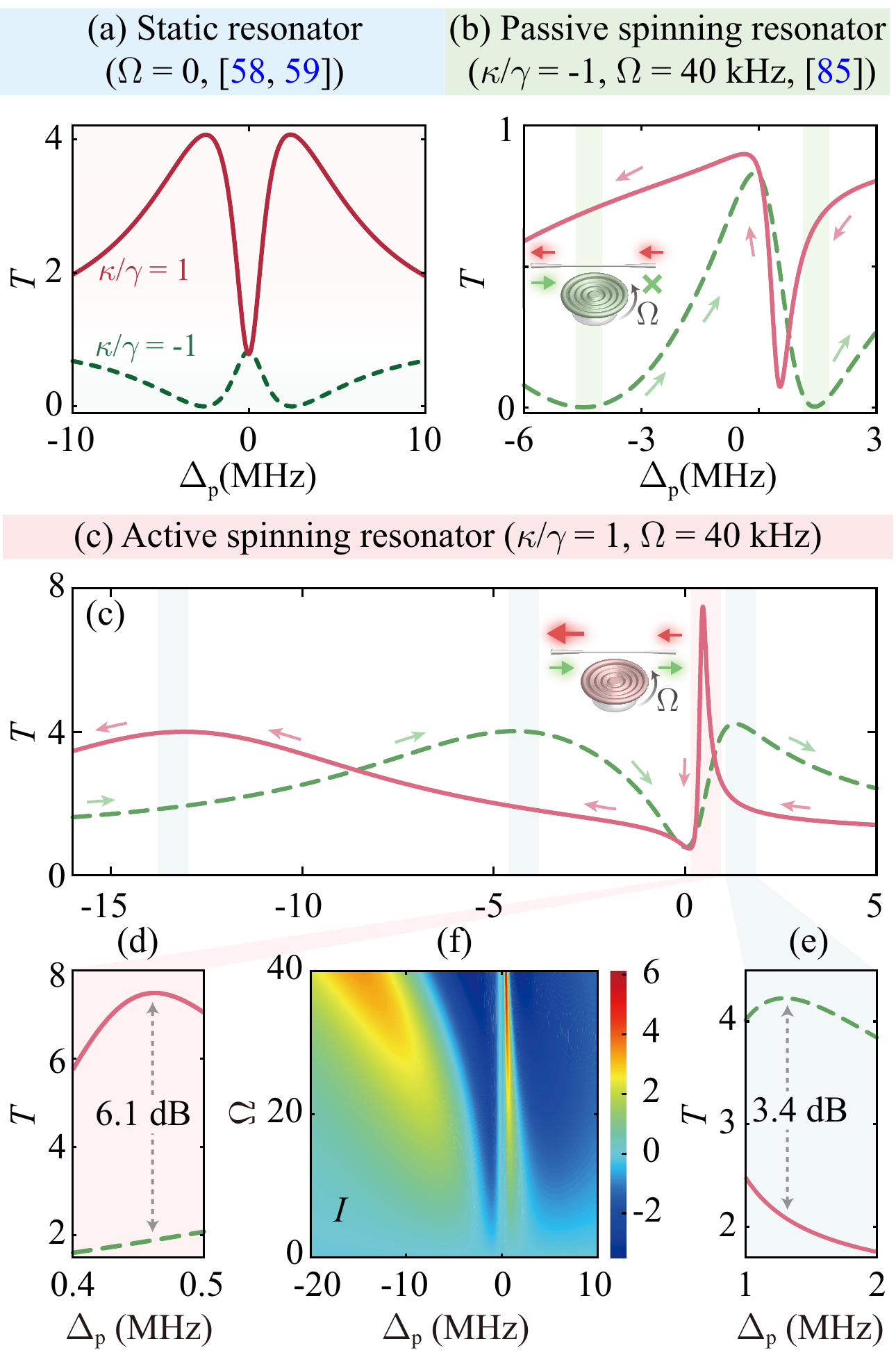}
		\caption{(a) Inverted-OMIT~\cite{jing2015optomechanically,jiao2016nonlinear}: Transmission rate $T$ of the probe field as a function of the detuning $\Delta_{\mathrm{p}}=\omega_{\mathrm{p}}-\omega_{\mathrm{c}}$ for different values of the gain-to-loss ratio $\kappa/\gamma$ in a static resonator. (b) Nonreciprocal transmission~\cite{lu2017optomechanically}: $T$ versus $\Delta_{\mathrm{p}}$ for different input directions in a passive spinning resonator. (c-e) Nonreciprocal amplification: $T$ versus $\Delta_{\mathrm{p}}$ for different input directions in an active spinning resonator. Arrows correspond to input on the left-hand or right-hand side. (f) Isolation ratio $I$ versus $\Delta_{\mathrm{p}}$ and the rotating speed $\Omega$ in the active resonator. The parameters are selected as listed in the table of Fig.~\ref{fig:Schematic diagram}(b).}\label{fig:transmission rate}
	\end{figure}

	In our numerical simulations, the experimentally accessible parameters are chosen, as shown in Fig.~\ref{fig:Schematic diagram}(b). According to a very related experiment~\cite{maayani2018flying}, a spherical cavity with a radius of $1.1\;\mathrm{mm}$ can steadily revolve around its axis at a speed of $6.6\;\mathrm{kHz}$. Theoretically, if the driving power is certain, the maximum speed is inversely proportional to the square of the radius; this means that the smaller the cavity, the greater the speed~\cite{jiang2022nonlinear}. Taking these factors into account, we use a rotary speed of no more than $40\;\mathrm{kHz}$.

	\section{Results and discussions}\label{Sec: Results and discussions}
	\subsection{The transmission rate of the probe field}
	In our work, we are interested in the role of both optical gain and mechanical rotations in the manipulation of the transmission spectrum. For comparison, we first consider the case of a static resonator. Figure~\ref{fig:transmission rate}(a) shows the transmission rate $T$ of the probe field versus detuning $\Delta_{\mathrm{p}}=\omega_{\mathrm{p}}-\omega_{\mathrm{c}}$ for different values of the gain-to-loss ratio $\kappa/\gamma$. In the standard COM system (i.e., a passive static COM resonator), a standard single transparency window emerges around the resonance point \cite{weis2010optomechanically} [see the green dashed curve in Fig.~\ref{fig:transmission rate}(a)]. This is also consistent with the description in Refs.~\cite{jing2015optomechanically,jiao2016nonlinear}, when the amount of gain provided to the resonator supersedes its loss, and the resonator becomes an active resonator, an inverted-OMIT profile appears for the probe field [see the red solid curve in Fig.~\ref{fig:transmission rate}(a)], which can be viewed as an analog of the optical inverted-EIT~\cite{oishi2013inverted}. Moreover, according to our analysis, when increasing the gain-to-loss ratio until it approaches to balance ($\kappa/\gamma=0$), the system tends to evolve into an unstable region.
	
	In addition, we have also confirmed that in a spinning COM system without any gain, the nonreciprocal transmission of light can be achieved, which is reminiscent of the result discussed in Ref.~\cite{lu2017optomechanically}. Figure~\ref{fig:transmission rate}(b) shows the transmission rate $T$ of the probe field versus optical probe detuning $\Delta_{\mathrm{p}}$ for different input directions in the passive spinning resonator. For a fixed rotary direction, the probe light coming from the left or the right can be blocked or transmitted due to the rotation-induced Sagnac frequency shift, leading to the emergence of nonreciprocal light propagation~\cite{lu2017optomechanically}.
	
	Inspired by these preceding works, we now study how to achieve a nonreciprocal optical amplifier with an active spinning optomechanical resonator. In Figs.~\ref{fig:transmission rate}(c)-(e), we depict the effect of rotation on the signal transmission rate in an active COM system. It is worth noting that, in the presence of rotation, a wider OMIT linewidth can be obtained by increasing the spinning speed, resulting in a Fano-like inverted-OMIT transparency spectrum [see Fig.~\ref{fig:transmission rate}(c)]. In particular, the amplification of the nonreciprocal signal is easily obtained by tuning  the directions of the input probe field, and the nonreciprocal isolation can be as high as $6\;\mathrm{dB}$ by selecting the appropriate speed and detuning [see Fig.~\ref{fig:transmission rate}(d)], which is caused by the interplay of the Sagnac effect and the optical gain. This essential physics can be understood as follows: as shown in Refs.~\cite{jing2015optomechanically,jiao2016nonlinear}, increasing the gain can turn a conventional OMIT profile into an inverted-OMIT profile. Meanwhile, the rotation of the resonator increases the amplitude of the steady-state mechanical displacement explosively. For example, when choosing $\Omega=40\;\mathrm{kHz}$, $x_{\mathrm{s}}$ can be increased from $80\;\mathrm{fm}$ to $10\;\mathrm{pm}$, thus leading to a strong enhancement of the effective COM coupling rate $G=gx_{\mathrm{s}}$~\cite{lu2017optomechanically}. Hence, when $\Omega$ and $\Delta_p$ are kept fixed, a nonreciprocal optical amplifier can be realized due to the interplay between the Sagnac effect and gain materials. Moreover, the multi-color nonreciprocal light amplification is observed by tuning the directions of the input driving field [see Figs.~\ref{fig:transmission rate}(c-e)].
	
	Depending on the direction of the input driving field, the transmittance of the probe field exhibits a highly asymmetric feature because the optical frequency shift caused by rotation is related to the driving direction~\cite{lu2017optomechanically}. In order to better understand and visualize the effect, we define the isolation ratio as~\cite{zhang2020breaking}
	\begin{align}
		I=10\mathrm{log}_{10}\frac{T\left(\Omega<0\right)}{T\left(\Omega>0\right)}.
	\end{align}Fixing the CCW rotation of the resonator, $\Omega>0$ ($\Omega<0$) denotes the driving field input from the left-hand (right-hand) side. Figure.~\ref{fig:transmission rate}(f) shows that the isolation ratio $I$ depends on the optical detuning $\Delta_{\mathrm{p}}$ for different values of the rotary speed $\Omega$ to give a comprehensive view. 	
\begin{figure}[ht]
	\centering
	\includegraphics[width=\linewidth,scale=0.95]{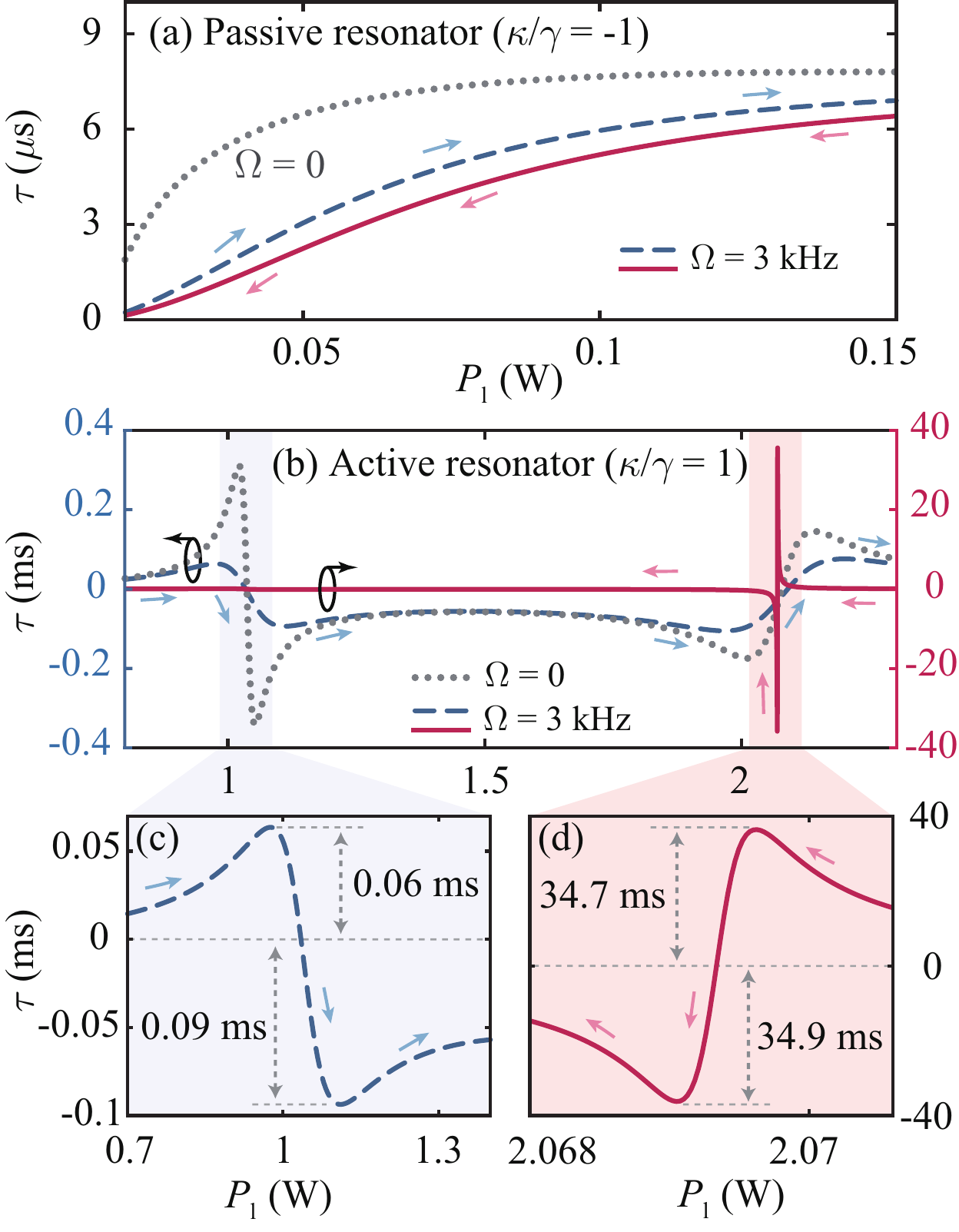}
	\caption{Group delay $\tau$ of the probe light as a function of the pump power $P_{\mathrm{l}}$ for different input directions in (a) a passive COM resonator or (b-d) an active COM resonator. Arrows correspond to input on the left-hand or right-hand side. Here  $\Omega=3\;\mathrm{kHz}$ for the spinning resonator, and $\Omega=0$ for the nonspinning resonator. For the other parameter values, see the table in Fig.~\ref{fig:Schematic diagram}(b).}\label{fig:group delay}
\end{figure}
The maximum value of isolation $I$ increases with a higher rotary speed in the active resonator. This implies that, by transforming $\Delta_{\mathrm{p}}$ and $\Omega$, the gain of light with direction-dependence can be controlled to achieve a unique nonreciprocal feature. It is obvious that the nonreciprocal amplification effect will be further enhanced, if we can continue to increase the speed of the active cavity.
	
	\subsection{Optical group delay}

	Accompanied with the OMIT, slowing or advancing of light can also be observed~\cite{safavi2011electromagnetically,weis2010optomechanically,jing2015optomechanically}. Specifically, the group delay of the probe light is given by
	\begin{align}\label{key1}
		\tau_{\mathrm{g}}=\left.\frac{d\,\mathrm{arg}\left(t_{\mathrm{p}}\right)}{d\,\xi}\right|_{\xi=\omega_{\mathrm{m}}}.
	\end{align}
	To see this, we plot the group delay of the probe light $\tau_{\mathrm{g}}$ as a function of the driving power $P_{\mathrm{l}}$ for different values of gain-to-loss ratio $\kappa/\gamma$ in Fig.~\ref{fig:group delay}. 	
	\begin{figure*}[t]
		\centering
		\includegraphics[width=0.95\textwidth]{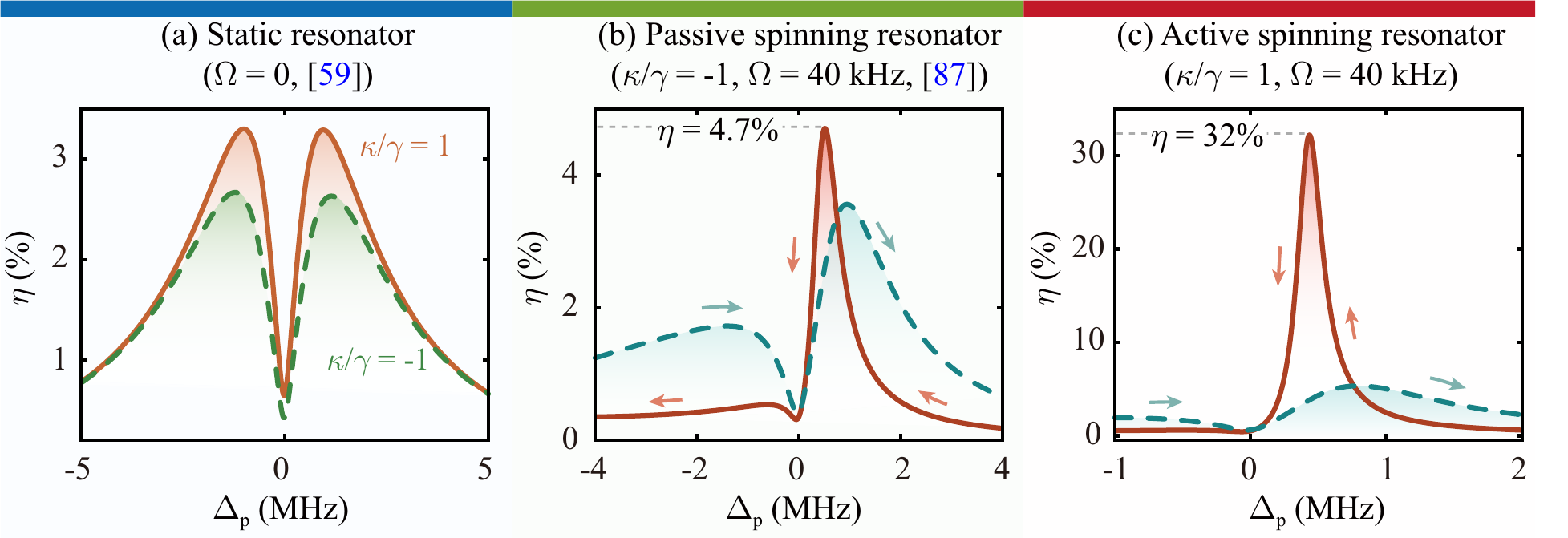}
		\caption{(a) The efficiency $\eta$ of the second-order sideband as a function of the optical detuning $\Delta_{\mathrm{p}}$ for different values of the gain-to-loss ratio $\kappa/\gamma$ in the static resonator~\cite{jiao2016nonlinear}. $\eta$ versus $\Delta_{\mathrm{p}}$ for different input directions in (b) the passive resonator~\cite{li2020nonreciprocal} or (c) the active resonator. Arrows correspond to input on the left or right side. The parameters are selected as listed in the table of Fig.~\ref{fig:Schematic diagram}(b).}\label{fig:Second-order sideband}
	\end{figure*}
	We find that the dispersion of the spinning active resonator can be severely affected by the interplay of the Sagnac effect and the optical gain~\cite{jing2015optomechanically,jiao2016nonlinear,lu2017optomechanically}, resulting in the rotary speed and the driving direction to become the two control knobs to regulate the group delay. For a passive spinning resonator, we have confirmed that OMIT leads only to the slowing of the transmitted light~\cite{lu2017optomechanically}. In contrast, in a static active resonator (i.e., $\kappa/\gamma=1$ and $\Omega=0$), one can tune the system to switch from slow to fast light, or vice versa, by tuning the driving power $P_{\mathrm{l}}$~\cite{jing2015optomechanically,jiao2016nonlinear}. Moreover, the group delay $\tau_{\mathrm{g}}$ can also be significantly enhanced by about 40 times for their maximum values in comparison with the passive static resonator~\cite{jing2015optomechanically,jiao2016nonlinear}. In particular, by spinning the active resonator (i.e., $\kappa/\gamma=1$ and $\Omega \neq0$), nonreciprocal group delay or advance can be realized [see Fig.~\ref{fig:group delay}(b)]. More importantly, in an active spinning resonator, the group delay $\tau_{\mathrm{g}}$ can be further significantly enhanced for the case with the driving field input from the right-hand side[see Figs.~\ref{fig:group delay}(b) and~\ref{fig:group delay}(d)]. For example, the maximum value of the group delay $\tau_{\mathrm{g}}$ is about $35\;\mathrm{ms}$ for $\Omega=-3\;\mathrm{kHz}$, i.e., about $111$ times enhancement compared to that for $\Omega=0$.

	\subsection{Nonlinear second-order sideband}

We note that nonlinear OMIT effects, usually exhibiting as weak signals of higher-order optical sidebands, can arise due to nonlinear COM interactions~\cite{xiong2012higher,Lemonde2013Nonlinear,liu2013Parametric,kronwald2013optomechanically}. The possibility of amplifying these weak signals has been explored in very recent works~\cite{jiao2016nonlinear,jiao2018optomechanical,li2016giant,suzuki2015nonlinear,lu2023magnon,Reynoso2022Optomechanical}, with a static active cavity or coupled resonators~\cite{jiao2016nonlinear}, a nonlinear Kerr resonator~\cite{jiao2018optomechanical}, a parity-time symmetric system consisting of a passive nonlinear cavity coupled to an active linear cavity~\cite{li2016giant}, a mechanically pumped COM system~\cite{suzuki2015nonlinear}, and a magnomechanical system~\cite{lu2023magnon}. Also, nonreciprocal enhancement of such sidebands can also be achieved by considering a spinning passive resonator~\cite{li2020nonreciprocal,wang2023nonreciprocal} or a nonlinear opto-magnonic system~\cite{wang2021nonreciprocal}. Here we show that well-controllable nonreciprocal amplifications of nonlinear OMIT effects can be realized in our present system.

To verify the role of optical gain and mechanical rotation in enhancing the second-order sideband in our system, we plot Figs.~\ref{fig:Second-order sideband}(a)-(c) to demonstrate that the efficiency $\eta$ varies with the optical detuning $\Delta_{\mathrm{p}}$. For a passive static resonator ($\Omega=0$ and $\kappa/\gamma=-1$), the second-order sideband is subdued when the OMIT emerges~\cite{xiong2012higher}, which results in a local minimum between the two sideband peaks around $\Delta_{\mathrm{p}}=0$ [see the green dashed curve in Fig.~\ref{fig:Second-order sideband}(a)]. As we know from Ref.~\cite{jiao2016nonlinear}, in an active static resonator, the double-peak spectrum still exists and $\eta$ is enhanced by the gain [see the red solid curve in Fig.~\ref{fig:Second-order sideband}(a)]. However, in a spinning COM system without any gain, the nonreciprocal enhancement of the second-order sideband can be achieved, which is reminiscent of the result discussed in Ref.~\cite{li2020nonreciprocal}. As shown in Fig.~\ref{fig:Second-order sideband}(b), when driving the system from the right-hand side, the left peak is completely suppressed and the right peak is only marginally enhanced compared to driving the system from the left-hand side. This result means that, by rotating the resonator, nonreciprocal enhancement of the second-order sideband can be achieved in a passive resonator due to the Sagnac effect~\cite{li2020nonreciprocal}.

More interestingly, for an active device, the second-order sideband can be further enhanced dramatically compared to the passive device. For example, when driving the system from the right-hand side, the efficiency $\eta$ is about $32\%$ for $\kappa/\gamma=1$ [see the red solid curve in Fig.~\ref{fig:Second-order sideband}(c)]., i.e., about 7 times enhancement compared to that for $\kappa/\gamma=-1$. The reason for the phenomenon is that the interplay of the Sagnac effect and the gain results in the left peak of the second-order sideband to be completely suppressed while the right peak is considerably enhanced.

\section{Conclusions}\label{Sec:Conclusions}
In conclusion, we have theoretically studied how to achieve a multi-color nonreciprocal optical amplifier by spinning an active optomechanical resonator. We find that in such an active spinning COM system, the optical Sagnac effect and the optical gain strongly impact the optical properties, including the signal transmission and its higher-order sidebands. As a result, nonreciprocal signal amplification can be realized, accompanied by a giant enhancement of the optical group delay from $0.3\;\mathrm{ms}$ to $35\;\mathrm{ms}$ in a chosen direction, which is useful for optical storage. Moreover, nonreciprocal enhancement of the second-order sideband can be further enhanced supernormally compared to the passive spinning resonator. These features of nonreciprocal optical amplification and giant enhancement of the optical group delay provide more flexible ways in practical applications ranging from multi-color optical communications to optical storage. Our work can also be extended to study enhancement of sensing, for instance, nanoparticle sensing~\cite{jing2018nanoparticle,sahin2014highly}, the gyroscope~\cite{zhang2022anti}, and nonreciprocal topological photonics~\cite{Zhang2022Topological}.
		
\begin{acknowledgements}
H.J. is supported by the National Natural Science Foundation of China (11935006) and the Science and Technology Innovation Program of Hunan Province (2020RC4047). T.-X. L. is supported by the NSFC (Grant No. 12205054), the Jiangxi Provincial Education Office Natural Science Fund Project (GJJ211437), and Ph.D. Research Foundation (BSJJ202122). Y.-F. J. is supported by the NSFC (Grant No. ~12147156), the China Postdoctoral Science Foundation (Grants No. ~2021M701176 and No.~2022T150208) and the Science and Technology Innovation Program of Hunan Province (Grant No.~ 2021RC2078).
\end{acknowledgements}
		

%

\end{document}